\documentclass[aps,prl,twocolumn,showpacs,superscriptaddress,floatfix]{revtex4}
\usepackage{epsfig}
\usepackage{amsmath}
\usepackage{bbm}

\newcommand{\bra}[1]{\langle #1|}
\newcommand{\ket}[1]{|#1\rangle}

\begin{document}

\title{Light-shift-induced photonic nonlinearities}

\author{F.G.S.L. Brand\~ao}
\affiliation{QOLS, Blackett Laboratory, Imperial College London, London SW7 2AZ, UK}
\affiliation{IMS, Imperial College London, 53 Prince's Gate Exhib.
Rd, London SW7 2PG, UK}
\author{M.J. Hartmann}
\affiliation{QOLS, Blackett Laboratory, Imperial College London, London SW7 2AZ, UK}
\affiliation{IMS, Imperial College London, 53 Prince's Gate Exhib.
Rd, London SW7 2PG, UK}
\author{M.B. Plenio}
\affiliation{QOLS, Blackett Laboratory, Imperial College London, London SW7 2AZ, UK}
\affiliation{IMS, Imperial College London, 53 Prince's Gate Exhib.
Rd, London SW7 2PG, UK}
\date{\today}

\begin{abstract}

We propose a new method to produce self- and cross-Kerr photonic nonlinearities, using light-induced Stark shifts due to the interaction of a cavity mode with atoms. The proposed experimental set-up is considerably simpler than in previous approaches, while the strength of the nonlinearity obtained with a single atom is the same as in the setting based on electromagnetically induced transparency. Furthermore our scheme can be applied to engineer effective photonic nonlinear interactions whose strength increases with the number of atoms coupled to the cavity mode, leading to photon-photon interactions several orders of magnitude larger than previously considered possible.       
     
\end{abstract}

 \maketitle

Quantum properties of light, such as photon anti-bunching \cite{walls} and photonic entanglement \cite{raimond}, can only be produced by nonlinear interactions between photons. Strong nonlinearities are also important for quantum information processing, with applications ranging from quantum nondemolition measurements \cite{imoto} to quantum memories for light \cite{lukin} and optical quantum computing architectures \cite{kimble}. However, photon-photon interactions are usually extremely weak, and several orders of magnitude smaller than those needed in the above applications. A possible route towards larger nonlinearities is the use of coherent interaction between light and matter in high finesse QED cavities \cite{mabuchi}. The goal here is to produce large nonlinearities with negligible losses, something which cannot be accomplished by merely tuning close to resonance atom-light interactions. To the best of our knowledge, the setting generating the largest nonlinear interactions in this context, proposed by Imamo\u{g}lu and co-workers \cite{imamoglu, hartmann1, Hartmann2}, uses quantum interference effects related to electromagnetically induced transparency (EIT) \cite{morangos} to produce giant Kerr nonlinearities, essentially absorption free, that are up to 9 orders of magnitude larger than natural Kerr interactions (see Fig 1. (a)) \footnote{As shown in Ref. \cite{hartmann1}, the EIT setting can produce nonlinearities with a strength of $\alpha \beta g$, where $g$ is the Rabi frequency of the atoms-cavity interaction and $\alpha, \beta$ are two parameters which must be much smaller than one.}. Another distinctive feature of this scheme is that the nonlinearity stays constant when several atoms interact with the same cavity mode. 

\begin{figure}
\begin{center}
\includegraphics[scale=0.4]{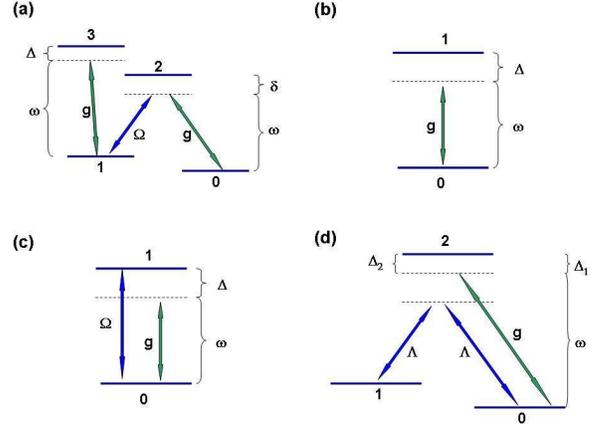}
\caption{(a) EIT scheme considered in Refs. \cite{imamoglu, hartmann1}, (b) dispersive regime of the Jaynes-Cummings interaction, (c) large but noise nonlinearity scheme, and (d) scheme proposed in this paper. Coupling to the cavity mode is shown in green and to classical lasers in blue.}
\end{center}
\end{figure}

In this Letter we propose a new method for producing Kerr nonlinearities in cavity QED systems which (i) is experimentally less demanding, (ii) virtually absorption free, and (iii) produces nonlinearities comparable or even superior to the state-of-art EIT scheme \cite{imamoglu}. By applying suitable laser pulses at the beginning and end of the evolution of the proposed set-up, we obtain nonlinear interactions whose (iv) strength increases with increasing number of atoms interacting with the cavity mode, leading to effective nonlinear interactions at least two orders of magnitude larger than previously considered possible. This brings closer to reality a number of proposals for quantum computation and communication based on photonic nonlinearities. In particular, the estimated strength of our nonlinearity is sufficiently large to allow the implementation of quantum nondemolition measurements \cite{munro0}, a photonic CNOT gate \cite{munro}, and continuous variable entanglement distillation \cite{cirac}. 

In our approach, the relevant atomic level structure, depicted in Fig. 1 (d), is a $\Lambda$ system with two metastable states and an excited state. The cavity mode couples dispersively only to the $0-2$ transition and the levels $0$ and $1$ are coupled via a far-detuned Raman transition. Finally, the detuning associated with the lasers and with the cavity mode are assumed to be very different from each other. This experimental setting can be implemented in several cavity QED systems, including atoms trapped in Fabry-P\'erot cavities \cite{raimond} or optical microcavities \cite{aoki, hinds} and quantum dots embedded in photonic band gap structures \cite{ima}. In particular, our proposal is of considerable importance for the realization of many-body phenomena in systems consisting of arrays of coupled cavities. There it was proposed \cite{Hartmann2, others} that quantum-phase transitions for polaritons and photons, as well as a photonic Mott-insulator \cite{hartmann1} - in which photons would be frozen inside each cavity, mimicking a light crystal - could be experimentally observed provided strong photonic nonlinearities are available.

To understand intuitively how our scheme works, let us start with the simplest system producing a nonlinearity: $N$ two level atoms interacting dispersively with a cavity mode $a$ (see Fig. 1 (b)). It is well known that if $\sqrt{N} g/ \Delta \ll 1$, where $g$ is the Rabi frequency of the Jaynes Cummings interaction and $\Delta$ the detuning, then the system is described, including fourth order terms in the perturbation theory and neglecting non-energy preserving terms, by the following Hamiltonian: 
\begin{eqnarray*}
H_{eff}^1 = \frac{N g^2}{\Delta} a^{\cal y}a (S_{00} - S_{11}) + \frac{N g^4}{\Delta^3} a^{\cal y}a a^{\cal y}a (S_{00} - S_{11}), 
\end{eqnarray*}
where $S_{aa} := \sum_{k=1}^N \ket{a_k}\bra{a_k}$. Hence, if all the atoms are prepared in their ground states, we obtain a self Kerr nonlinearity of strength $Ng^4/\Delta^3$. Despite the simplicity of this scheme, it has two major drawbacks. First, the obtained nonlinearities are at least one order of magnitude smaller than in the EIT setting [20]. Second, due to the adiabatic condition $\sqrt{N} g/ \Delta \ll 1$ characterizing the dispersive regime, the nonlinearity decreases as $g/\sqrt{N}$.  

A simple solution to both these problems is to add a classical laser in resonance to the transition $\ket{0} \rightarrow \ket{1}$ (see Fig. 1 (c)). Under the conditions $\sqrt{N} g/ \Delta \ll 1$ and $\sqrt{N} g^2 / \Delta \ll \Omega$, the dynamics of the system is well described by
\begin{equation} \label{h1}
H_{eff}^2 = \frac{N g^2}{\Delta} a^{\cal y}a S_{3} + \underbrace{\frac{\sqrt{N} g}{\Delta \Omega} }_{\ll 1} \underbrace{ \frac{ \sqrt{N}g^2 }{\Delta \Omega} }_{\ll 1} g a^{\cal y}a a^{\cal y}a S_{3},
\end{equation}
where $S_{3} := \sum_{k=1}^N \ket{+_k}\bra{+_k} - \ket{-_k}\bra{-_k}$, with $\ket{\pm} = (\ket{0} \pm \ket{1})/\sqrt{2}$. Hence, preparing the atomic states in the superposition $\ket{-}$, we find a nonlinearity as large as in the EIT setting [20] which does not decrease with the number of atoms. Unfortunately, all this improvement comes at the price of a prohibitively large increase in the losses of the system. Spontaneous emission will destroy the atomic superpositions, quickly decreasing the effective nonlinearity and introducing noise to the system.  

In the scheme that we propose (see Fig. 1 (d)), we have a situation similar to the one discussed in the previous paragraph. Indeed, the effective Hamiltonian will be identical to Eq. (\ref{h1}). However, now the states $\ket{\pm}$ are metastable, which makes the system almost decoherence free. The full Hamiltonian of the system, in the interaction picture, reads:
\begin{eqnarray} \label{hfull}
H &=& g \left( e^{- i \Delta_1 t} a S_{20} + e^{ i \Delta_1 t} a^{\cal y} S_{02} \right)  \\ &+& \sqrt{2} \Lambda \left( e^{- i \Delta_2 t} S_{2+} +  e^{- i \Delta_2 t} S_{+2}\right), \nonumber
\end{eqnarray}
where $S_{20} := \sum_k \ket{2_k}\bra{0_k}$ and  $S_{2+} := \sum_k \ket{2_k}\bra{+_k}$. As shown in Fig. 1 (d), ($g$, $\Delta_1$) and $(\Lambda$, $\Delta_2$) are the Rabi frequencies and detunings of the cavity-atom and laser-atom interactions, respectively. We are interested in the dispersive regime, characterized by:
\begin{equation} \label{cond3}
\frac{\sqrt{N} g}{\Delta_1} \ll 1, \hspace{0.1 cm} \frac{ \sqrt{N} \Lambda}{\Delta_2} \ll 1.
\end{equation}
Moreover, we assume that
\begin{equation}
\sqrt{N}g, \sqrt{N} \Lambda \ll |\Delta_2 - \Delta_1|,
\end{equation}
so that we can treat the processes driven by the cavity-atom and laser-atom interactions independently. Then, the excited state will hardly be populated, and we can adiabatically eliminate it, finding an effective Hamiltonian for the two metastable states and the cavity mode. In turn, these will experience a.c. Stark shifts due to the interaction with the upper level. The effective Hamiltonian, dropping out terms proportional to the identity, is given by 
\begin{equation}
H_1 = \frac{g^2}{\Delta_1}a^{\cal y}a S_{00} + \frac{\Theta}{2} (S_{10} + S_{01}),
\end{equation}
with $\Theta := 2\Lambda^2/\Delta_2$. If we now go to a second interaction picture with respect to $H_0 = \frac{g^2}{2\Delta_1}a^{\cal y}a$ we find
\begin{equation} \label{h3}
H_1^{int} = \frac{g^2}{2\Delta_1} a^{\cal y}a \left( S_{+-} + S_{-+}\right) + \frac{\Theta}{2} S_3 ,
\end{equation}
where $S_{+-} := \sum_k \ket{+_k}\bra{-_k}$, $S_{-+} = (S_{+-})^{\cal y}$, and, as before, $S_3 := \sum_{k=1}^N \ket{+_k}\bra{+_k} - \ket{-_k}\bra{-_k}$. It follows that in the $\{ \ket{\pm} \}$ basis the system can be viewed as an ensemble of two level atoms driven by a laser with a photon-number-dependent Rabi frequency. If we consider the dispersive regime of this system, i.e.
\begin{equation} \label{cond1}
\frac{\sqrt{N}g^2}{2\Delta_1 \Theta} \ll 1,
\end{equation}
the atoms prepared in the $\ket{-}$ state will experience a Stark shift proportional to $(a^{\cal y}a)^2$, which gives rise to the desired Kerr nonlinearity. The effective Hamiltonian will be given by $H_{eff} = \frac{g^4}{4\Delta_1^2\Theta} a^{\cal y}a a^{\cal y}a S_3$. Therefore, if we prepare all the atomic states in the $\ket{-}$ state, one obtains an essentially absorption free Kerr nonlinearity \footnote{As we performed two adiabatic eliminations, some higher order terms that we neglected in the first could be comparable with the ones we keep after the second elimination. An expansion to fourth order in the first elimination shows that all these terms lead to corrections of the form $\mu a^{\cal y}a \ket{\pm}\bra{\pm}$. Although these terms do not jeopardize in any sense the final effective model, they can be suppressed if (i) we impose the further condition $\frac{g^2}{\Delta_1} \gg \frac{\Lambda^3}{\Delta_2^2}$ and (ii) prepare all the atoms in the $\ket{-}$ state. } given by
\begin{equation} \label{h2}
H_{kerr} = \underbrace{\frac{\sqrt{N}g^2}{2\Delta_1 \Theta}}_{\ll 1} \underbrace{\frac{\sqrt{N} g}{2 \Delta_1}}_{\ll 1} g a^{\cal y}a a^{\cal y}a.
\end{equation}
From Eq. (\ref{h2}) we can see that the strength of the effective Kerr nonlinearity does not depend on the number of atoms, as we can tune $\Delta_1$ and $\Theta$ independently. As such, the maximum achievable nonlinearity is limited by $g$, which depends on the properties of the cavity and of the atoms employed. 

We now proceed to show that, in fact, our set-up can be modified to give nonlinear interactions which increase with $N$. The joint atomic operators $S_{+-}, S_{-+},$ and $S_3$ satisfy the $su(2)$ commutation relations: $[S_{+-}, S_{-+}] = S_3, [S_3, S_{\pm \mp}] = \pm S_{\pm \mp}$. Defining the canonical transformation $U = \exp(\mu a^{\cal y }a(S_{+-} - S_{-+}))$, with $\mu := g^2/(\Delta_1 \Theta) \ll 1$ \cite{ss}, we can use Hausdorff expansion ($\exp(x A)B \exp(-x A) = B + x[A, B] + x^2[A, [A, B]]/2 + ...$) to obtain from Eq. (\ref{h3})
\begin{eqnarray} \label{h5}
H_{rot} &:=& U^{\cal y}(H_1^{int})U \approx \frac{\Theta}{2} S_3 -  \left(\frac{g^2}{2\Delta_1}\right)^2 \frac{1}{\Theta} (a^{\cal y}a)^2S_3 \nonumber \\ &+& \left(\frac{g^2}{2\Delta_1}\right)^3 \frac{1}{\Theta^2} (a^{\cal y}a)^3(S_{+-} + S_{-+}). 
\end{eqnarray} 
Suppose we had a way of generating $H_{rot}$. If we prepared all the atoms in the $\ket{-}$ state, the effective photonic Hamiltonian would be given by the second term in Eq. (\ref{h5}) as long as
\begin{equation} \label{cond2}
\left(\frac{g^2}{2\Delta_1}\right)^3 \frac{\sqrt{N}}{\Theta^3} \ll 1.
\end{equation}
This condition ensures that basically no transition from $\ket{-}$ to $\ket{+}$ happens. Note that it is much less stringent than the one given by Eq. (\ref{cond1}). In particular, setting $\Theta$ such that $(g^2/2\Delta_1)/\Theta = N^{-1/4}$, Eq. (\ref{cond2}) is satisfied for large $N$, and we obtain a nonlinearity of $(\sqrt{N}g/\Delta_1) N^{1/4} g a^{\cal y}a a^{\cal y}a$. For instance, with $N = 10^4$ and $(\sqrt{N}g/\Delta_1) = 0.1$ , we obtain a nonlinearity equal to the Rabi frequency $g$, which is at least two orders of magnitude larger than in the single atom case.  

\begin{figure}
\begin{center}
\includegraphics[scale=0.4]{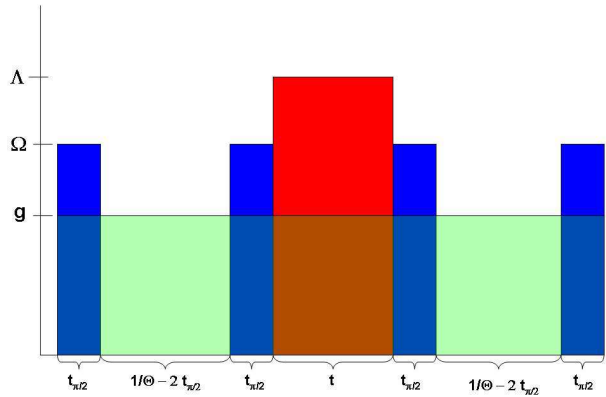}
\caption{Sequence of operations generating $V(t)$. Blue and red areas correspond to the time the lasers with Rabi frequencies $\Omega$ and $\Lambda$ are on, respectively. The green area illustrates that the Jaynes-Cummings interaction between the cavity mode and the atoms is on at all times. The scales do not reproduce reality in general. For example, $t$ will be much larger than $1/\Theta$ in most cases.}
\end{center}
\end{figure}
It is possible to realize the unitary operator $V(t) = \exp(i t H_{rot})$ for every $t$ (see Fig. 2). We have that $V(t) := U^{\cal y}\exp(i t H_1^{int})U$, hence it sufficies to show how to create the unitary $U$. Consider the Hamiltonian given by Eq. (\ref{hfull}) when the classical lasers are swiched off, i.e. $H = g(e^{-i \Delta_1 t}aS_{20} + e^{i \Delta_1 t}a^{\cal y}S_{02})$. Suppose we apply to the atoms a fast laser pulse, described by the unitary operator $M$ that generates the transformation
\begin{eqnarray*}
\ket{0_k} \rightarrow (\ket{0_k} + i\ket{1_k})/\sqrt{2} \\
 \ket{1_k} \rightarrow (\ket{0_k} - i\ket{1_k})/\sqrt{2},
\end{eqnarray*}
let Hamiltonian $H$ run for a time $t$ and apply the inverse transformation $M^{\cal y}$. Then, the total evolution operator will be given by
\begin{eqnarray*}
&&M^{\cal y} \exp(i g \int_0^t (e^{-i \Delta_1 t'}aS_{20} + e^{i \Delta_1 t'}a^{\cal y}S_{02})dt') M \\ &=& \exp(i g \int_0^t \sqrt{2}g(e^{-i \Delta_1 t'}a(S_{20} + i S_{21}) + h.c.)dt').
\end{eqnarray*}
As $\sqrt{N}g/\Delta_1 \ll 1$, the unitary operator above is well approximated by $\exp( t (g^2/\Delta_1)a^{\cal y}a( S_+ - S_-))$, which is $U$ when $t = 1/\Theta$. The sequence of operations executing $V(t)$ is shown in Fig. 2. 
\begin{figure}
\begin{center}
\includegraphics[scale=0.3]{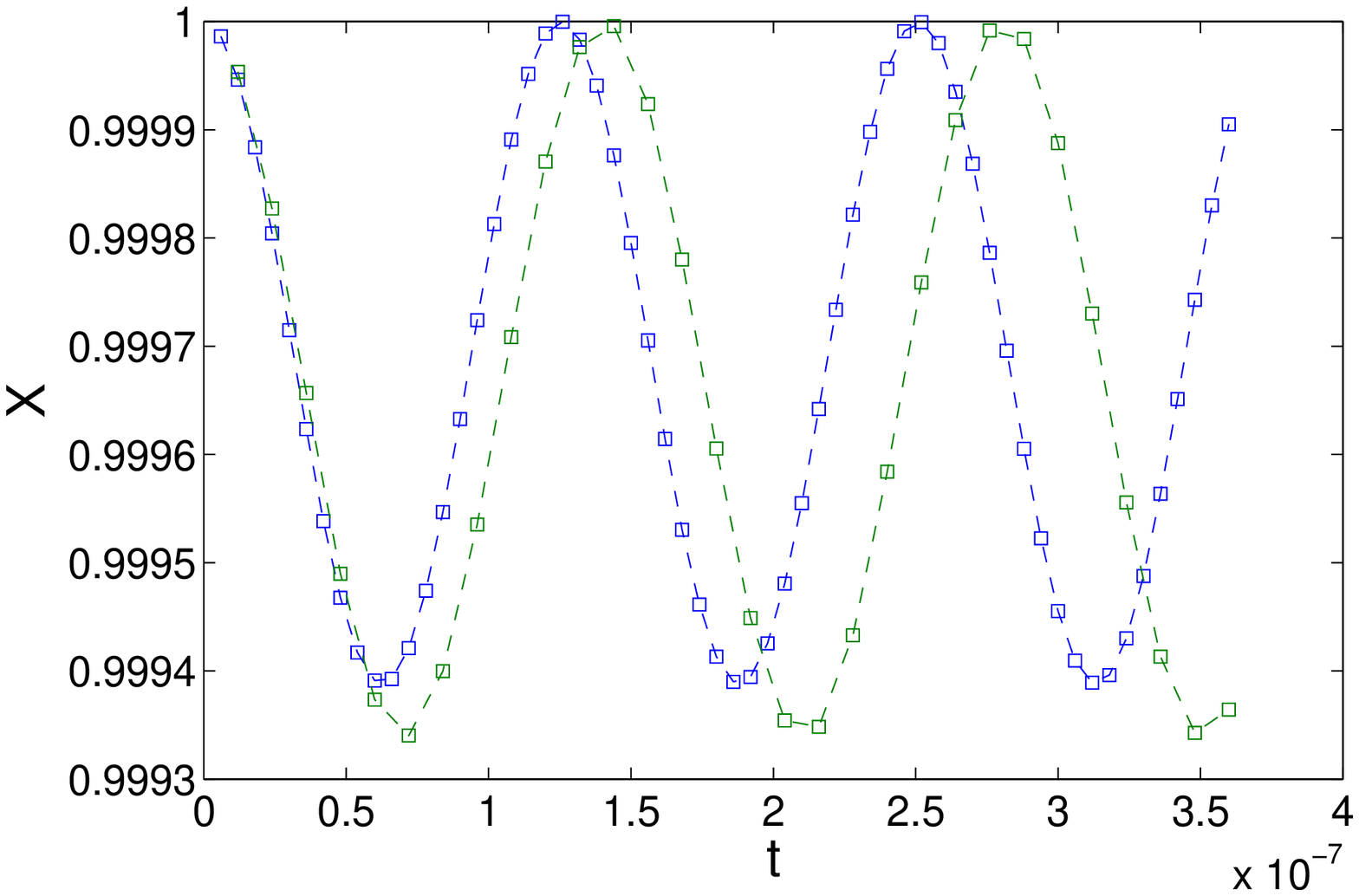}\hspace{-0.2cm}%
\includegraphics[scale=0.3]{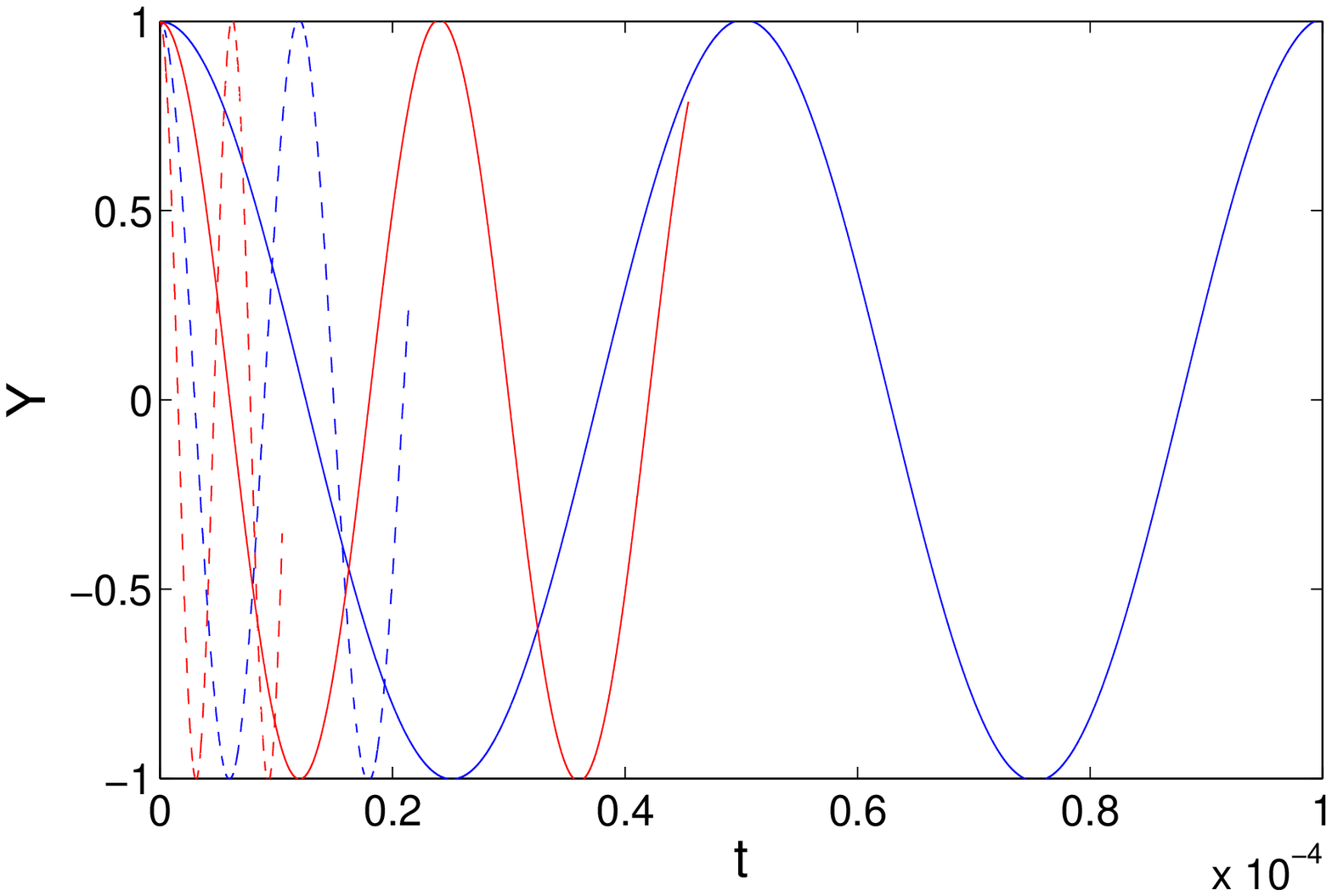}
\caption{(a) $X = |\bra{n} \otimes \bra{-}V(t)\ket{-}\otimes \ket{n}|$ versus $t$ for $(N, n) = (1, 2)$ (blue) and $(2, 2)$ (green), with $g = 10^8 s^(-1)$, $5 \Theta = gN^{1/3}$, $\Omega = 100g$, and $\Delta_1 = 10 \sqrt{N} g$; (b) $Y = Re(\bra{n} \otimes \bra{-}V(t)\ket{-}\otimes \ket{n})$ versus $t$ for $(N, n) = (1, 1)$ (blue solid line), $(1, 2)$ (blue dashed line), $(2, 1)$ (red solid line), and $(2, 2)$ (red dashed line), with $g = 10^8 s^{-1}$, $\Delta_1 = 10g$, $\Theta = g$, and $\Omega = 100g$.}
\end{center}
\end{figure}

In order to check the accuracy of this approach, we simulated it numerically for one and two atoms. We assumed that the laser pulse is generated by a $\pi/2$ pulse of the Hamiltonian $H_{laser} = \Omega (\ket{0}\bra{1} + \ket{1}\bra{0})$. We chosed $\Omega = 100g$, which gives a $t_{\pi/2} = \pi/(2\Omega)$. The other parameters are $g = 10^8 s^{-1}$, $5 \Theta = gN^{1/3}$, and $\Delta_1 = 10 \sqrt{N} g$. With these values, the quality of the approximation, which is determined by the value of the L.H.S. of Eq. (\ref{cond2})  and  $\sqrt{N}g/\Delta_1$, should be constant. A good figure of merit in this respect is given by $|\bra{n} \otimes \bra{-}V(t)\ket{-}\otimes \ket{n}|$, where $V(t)$ is the total unitary operator formed by concatenating the steps explained above (see Fig. 2 for a graphic description of the procedure), $\ket{n}$ is the $n$-th Fock state, and $\ket{-} := \ket{-_1}\otimes ... \otimes \ket{-_N}$ (see Fig. 3 (a)). In Fig. 3 (b) we plotted $Re(\bra{n} \otimes \bra{-}V(t)\ket{-}\otimes \ket{n})$ for $(N, n) = (1, 1), (1, 2), (2, 1), (2, 2)$, with the parameters $g = 10^8 s^{-1}$, $\Delta_1 = 10g$, $\Theta = g$, and $\Omega = 100g$. We found a very good agreement with the dynamics of a pure photonic nonlinearity of strenght $\kappa = N g^4 / 4\Delta_1^2 \Theta$, for which $Re(\bra{n} \otimes \bra{-}V(t)\ket{-}\otimes \ket{n}) = \cos(\kappa n^2 t)$.   

Up to now we have discussed the case of self-Kerr nonlinearities. In a cross-Kerr nonlinearity, one optical field induces a Stark shift in another field proportional to the intensity of the former. In terms of two optical modes $a$ and $b$, the Hamiltonian is given by $H_{cross} = \nu a^{\cal y}a b^{\cal y}b$. There are two possible generalizations of the scheme introduced here to produce large cross-Kerr nonlinearities. 

A shown in Fig. 4 (a), we may chose setting as in Eq. (\ref{hfull}), but now, we have a second cavity mode $b$ coupled to the transition $\ket{1} \rightarrow \ket{2}$ with Rabi frequency $g_b$ and Detuning $\Delta_{1_b}$. As shown in Ref. \cite{duan}, this could be achieved using two degenerate cavity modes with orthogonal polarization. Carrying over the same analysis we did for the one mode case, we can find in this case 
\begin{equation} 
H_{eff} = \frac{1}{\Theta}\left( \frac{g_a^2}{2 \Delta_{1_a}}a^{\cal y}a - \frac{g_b^2}{2 \Delta_{1_b}}b^{\cal y}b    \right)^2 S_3.
\end{equation}
Hence, in addition to a self-Kerr nonlinearity for each mode, we find a cross-Kerr nonlinearity between them. 
\begin{figure}
\begin{center}
\includegraphics[scale=0.4]{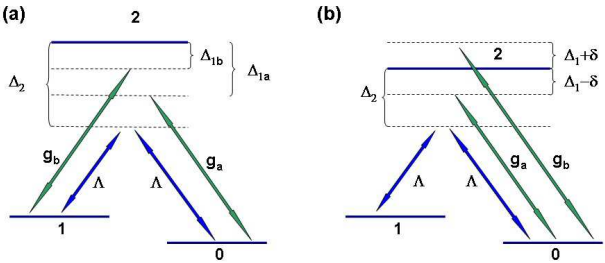}
\caption{Level structures for generating cross-Kerr nonlinearities.}
\end{center}
\end{figure}

In the second case, we consider the level structure shown in Fig. 4 (b). We have two modes $a$ and $b$ coupled to the transition $\ket{0} \rightarrow \ket{2}$ with detunings $\Delta_1 \pm \delta$, respectively. As discussed in Ref. \cite{aoki}, this set-up could be realized in a toroidal microcavity, where $a$ and $b$ are the normal modes of the clock and counter-clockwise propagating modes, and $\delta$ is the rate of tunneling between those two. In this case, the effective Hamiltonian in this case can be found to be
\begin{equation} 
H_{eff} = \frac{1}{\Theta}\left( \frac{g_a^2}{2 (\Delta_1 - \delta)}a^{\cal y}a + \frac{g_b^2}{2 (\Delta_1 + \delta)}b^{\cal y}b \right)^2 S_3
\end{equation}

Summarizing, we have presented a new proposal for producing giant Kerr nonlinearities in cavity QED systems. Our approach could be used to generate nonlinearities at least two orders of magnitude larger than previously considered possible. These, in turn, could be applied to the implementation of several quantum infomation processing tasks with photons, as well as paving a way to the observation of quantum many-body phenomena in coupled array of cavities. 

\textit{Acknowledgements}: This work is part of the QIP-IRC supported by EPSRC (GR/1582176/01), the EU integrated project QAP supported by the IST directorate under contract no. 015848 and was supported by the CNPq and a Royal Society Wolfson Research Merit Award and the Alexander von Humboldt Foundation.

\end{document}